\documentclass[sort&compress,10pt,twocolumn,final,3p]{elsarticle}

\usepackage{textcomp}
\usepackage{fleqn}
\usepackage{graphicx}
\usepackage{hyperref}
\usepackage{pifont}
\usepackage{amsmath,amssymb}
\usepackage{color,soul}
\usepackage{natbib}
\usepackage{miller}
\usepackage[utf8]{inputenc}
\usepackage{xspace}

\newcommand{\etal}{\textit{et al}.\@\xspace}
\newcommand{\abinitio}{\textit{ab initio}\@\xspace}
\newcommand{\Abinitio}{\textit{Ab initio}\@\xspace}

\begin{document}

\title{Basal slip of $\hkl<a>$ screw dislocations in hexagonal titanium}

\author[1]{Piotr Kwasniak\corref{cor1}}
\ead{piotr.kwasniak@pw.edu.pl}
\author[2]{Emmanuel Clouet}
\address[1]{Faculty of Materials Science and Engineering, Warsaw University of Technology, Woloska 141, 02-507 Warsaw, Poland}
\address[2]{DEN-Service de Recherches de Métallurgie Physique, CEA, Université Paris-Saclay, F-91191 Gif-sur-Yvette, France}
\cortext[cor1]{Corresponding author}

\begin{abstract} 
	Basal slip of $\hkl<a>$ screw dislocations in hexagonal closed-packed titanium 
	is investigated with \textit{ab initio} calculations.
	We show that a basal dissociation is highly unstable and reconfigures to other structures 
	dissociated in a first order pyramidal plane.
	The obtained mechanism for basal slip corresponds to the migration of the partial dislocations
	and of the associated stacking fault ribbon in a direction perpendicular
	to the dissociation plane.
	Presented results indicate that both basal and pyramidal slip will operate through 
	the Peierls mechanism of double-kink nucleation 
	and will be equally active at high enough temperature.
	
\end{abstract}

\begin{keyword}
Titanium; Plasticity; Dislocations; \textit{Ab initio}; Basal slip
\end{keyword}

\maketitle

Plastic deformation of hexagonal titanium depends on complex interplay of temperature, purity and activity of the different slip modes \cite{Churchman1954,Conrad1981,Naka1982,Caillard2003}. 
Experiments recognized that prismatic slip of $\hkl<a>=1/3\,\hkl<1-210>$ dislocations is the easiest slip mode with a lattice friction opposing the motion of the screw orientation for temperatures below 550\,K while $\hkl<a>$ dislocation with an edge character remain highly mobile \cite{Naka1988,Biget1989}. 
In situ TEM straining experiments \cite{Farenc1993} confirmed that deformation is governed at low temperatures by long rectilinear screw dislocations which glide in the \hkl{10-10} prismatic planes through a jerky motion associated with some cross-slip in the \hkl{10-11} pyramidal planes. 
This thermally activated mobility of the screw dislocation in pure Ti could be explained by a locking-unlocking mechanism \cite{Farenc1993,Farenc1995}: the screw dislocation, which is sessile in its ground state, needs to reconfigure to a higher energy configuration corresponding to a dissociation in a prismatic plane where the dislocation can easily glide. 
Various cores have been proposed for this ground state \cite{Sob1974,Sob1975,Naka1988}
and recent \abinitio calculations \cite{Clouet2015} have shown that it corresponds to a planar dissociation in the pyramidal plane.

Besides prismatic and pyramidal glide, $\hkl<a>$ dislocations in hcp Ti are also known 
to glide in the \hkl(0001) basal planes \cite{Levine1966,Shechtman1973,Akhtar1975}.
Recent TEM observations \cite{Caillard2018,Barkia2017} at and above room temperature
show straight screw dislocations moving viscously in the basal planes. 
This deformation mode is, however, part of profuse cross slip 
and is always linked with prismatic glide, leading to wavy slip traces. 
Caillard \etal \cite{Caillard2018} have proposed that basal glide 
of the screw dislocations proceeds through a kink-pair mechanism, 
where the screw dislocation remains dissociated either in a pyramidal or prismatic plane
and a kink-pair nucleates in the neighboring Peierls valley,
with the kinks lying in the basal plane.
A basal dissociation of the gliding dislocation will therefore not be required 
to allow for basal slip.
\Abinitio calculations have actually not managed to stabilize such a basal configuration 
of the screw dislocation until now \cite{Tarrat2014}, 
probably because the basal $I_2$ stacking fault associated with such a dissociation 
has a too high energy in titanium \cite{Kwasniak2014,Kwasniak2016,Yin2017a,Rodney2017}.
Here, we use \abinitio calculations to examine more closely the stability 
of such a basal dissociation of the $\hkl<a>$ screw dislocation in pure Ti,
before determining the transition pathway corresponding to basal slip,
taking into account all configurations which have been found stable 
for the $\hkl<a>$ screw dislocation.

\begin{figure*}[ht]
\centering
\includegraphics[width=1\textwidth]{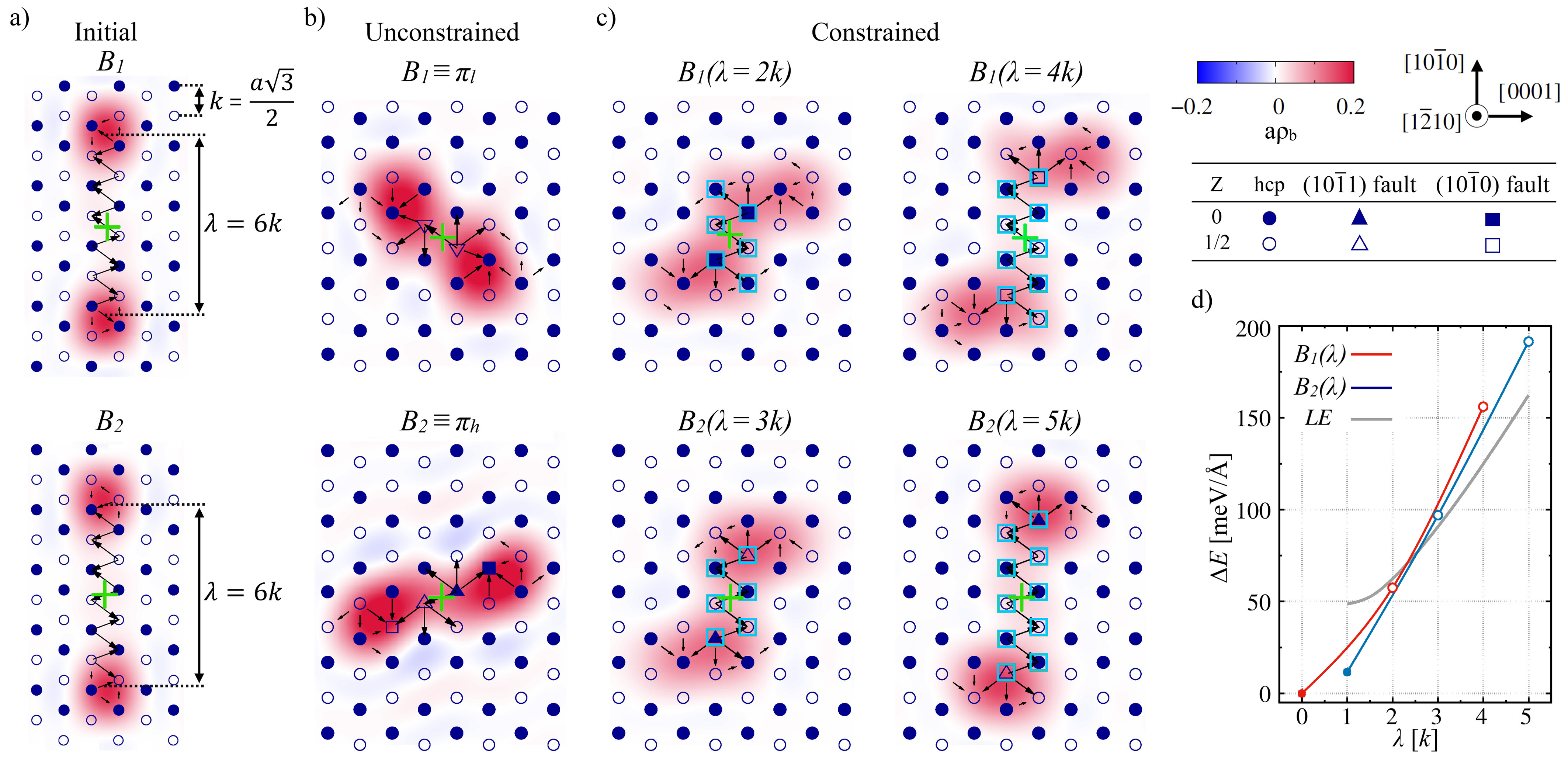}
\caption{Core structures corresponding to a basal dissociation of a $1/3\,\hkl[1-210]$ screw dislocation for two different  positions $B_1$ and $B_2$ of the dislocation center.
The initial configurations are shown in a),
the fully relaxed configurations in b)
and the configurations obtained with constrained minimization in c). 
Different dissociation distances $\lambda$ are studied.
The arrows between two atomic columns are proportional to the differential displacement created by the dislocation in the [1$\bar{2}$10] direction. 
Displacement smaller than 0.1\,\textit{b} are not shown. The contour map shows the dislocation density according to the Nye tensor. 
Ti atoms belonging to particular types of stacking faults or different (1$\bar{2}$10) atomic planes are plotted with open or full-colored symbols as presented in top-right corner. 
Positions of the dislocation centers are marked by green crosses and constrained atoms with a fixed position along the dislocation line are marked by light-blue squares. 
The energy variation $\Delta E$ as a function of the basal dissociation distance $\lambda$ obtained from \abinitio calculations ($B_1$ and $B_2$) and linear elasticity ($LE$) are shown in d) with filled and opened symbols respectively for unconstrained and constrained relaxations.}
\label{Fig1}
\end{figure*}

The computational supercell used in this study contains 288 atoms and exhibits \textit{m} = 9, \textit{n} = 8 and \textit{l} = 1 unit cell periodicity along $\vec{e\textsubscript{x}}$ = [0001], $\vec{e\textsubscript{y}}$ = [10$\bar{1}$0] and $\vec{e\textsubscript{z}}$ = 1/3[1$\bar{2}$10] directions, respectively. As a result, the periodicity vectors of the simulation box are $\vec{u\textsubscript{1}}$ = \textit{mc}$\vec{e\textsubscript{x}}$, $\vec{u\textsubscript{2}}$ = \textit{na}$\vec{e\textsubscript{y}}$ and $\vec{u\textsubscript{3}}$ = \textit{la}$\vec{e\textsubscript{z}}$, where \textit{a} and \textit{c} are the Ti lattice parameters. The calculations of the screw dislocations were performed in accordance to full periodic conditions approach \cite{Clouet2012,Rodney2017} and the initial structures of two line defects with opposite Burgers vectors were generated using anisotropic linear elasticity 
thanks to the \textsc{Babel} code \cite{Babel}.
The two dislocations composing the dipole are separated by a vector $\vec{d}=1/2\,(\vec{u}_1+\vec{u}_2)$,
thus leading to a quadrupolar arrangement which minimizes the elastic interaction between the dislocations and their periodic images.
The plastic strain created by the dipole is compensated by adding the tilt vectors $\vec{\delta\textsubscript{1}}$ = $\vec{b}$/2 = \textit{a}$\vec{e\textsubscript{z}}$/2 and $\vec{\delta\textsubscript{2}}$ = -$\vec{\delta\textsubscript{1}}$ to $\vec{u\textsubscript{1}}$ and $\vec{u\textsubscript{2}}$, respectively. 
It should be highlighted that the selected supercell geometry has an odd number of $\alpha$-Ti unit cell repetition along $\vec{e\textsubscript{x}}$ (\textit{m} = 9) which is different from those used in previous studies of \textlangle{}a\textrangle{} type screw dislocations in hcp systems \cite{Clouet2012,Chaari2014,Clouet2015,Kwasniak2017}.
This ensures that the plastic strain arising from the edge components of a basal dissociation
exactly compensates for the two dislocations composing the dipole,
thus preventing the existence of a back-stress caused by the fixed periodicity vectors
and which would acts against such a dissociation.

\mbox{\textit{Ab initio}} calculations were performed with VASP code \cite{Kresse1993,Kresse1996}, with projector augmented wave (PAW) method for core-valence electron interaction \cite{Blchl1994} and Perdew-Burke-Ernzerhof (PBE) \cite{Perdew2016} generalized gradient functional. 
The Brillouin zone was sampled in accordance with the \linebreak Monkhorst-Pack scheme \cite{Monkhorst1976} using $1\times1\times9$ gamma centered k-points grid and a 0.3\,eV Methfessel-Paxton electronic occupancy smearing. 
The Ti \textit{pv} pseudopotential with 10 valence electrons
($3p^6 3d^2 4s^2$) was employed, with a 500\,eV cutoff energy for plane waves.
The atomic structures of the screw dislocations were determined by performing 
ionic relaxations with a fixed-shape simulation box 
using a criterion of 3\,meV/\AA{} for Hellmann-Feynman forces convergence. 
The relative line defects energies are equal to:\\
\begin{equation}
\Delta{}E = \frac{E_{tot}-E_{\pi{}_l}}{2h}
\end{equation}
where, \textit{E\textsubscript{tot}} is the total energy of $\alpha$-Ti supercells with different positions or configurations of dislocation cores, \textit{E\textsubscript{$\pi$l}} is the total energy of the simulation box with dislocations in their ground state (low energy pyramidal configuration \cite{Clouet2015}), and $h$ is the height of the atomic model along dislocation line.

\begin{figure*}[ht]
\centering
\includegraphics[width=1\textwidth]{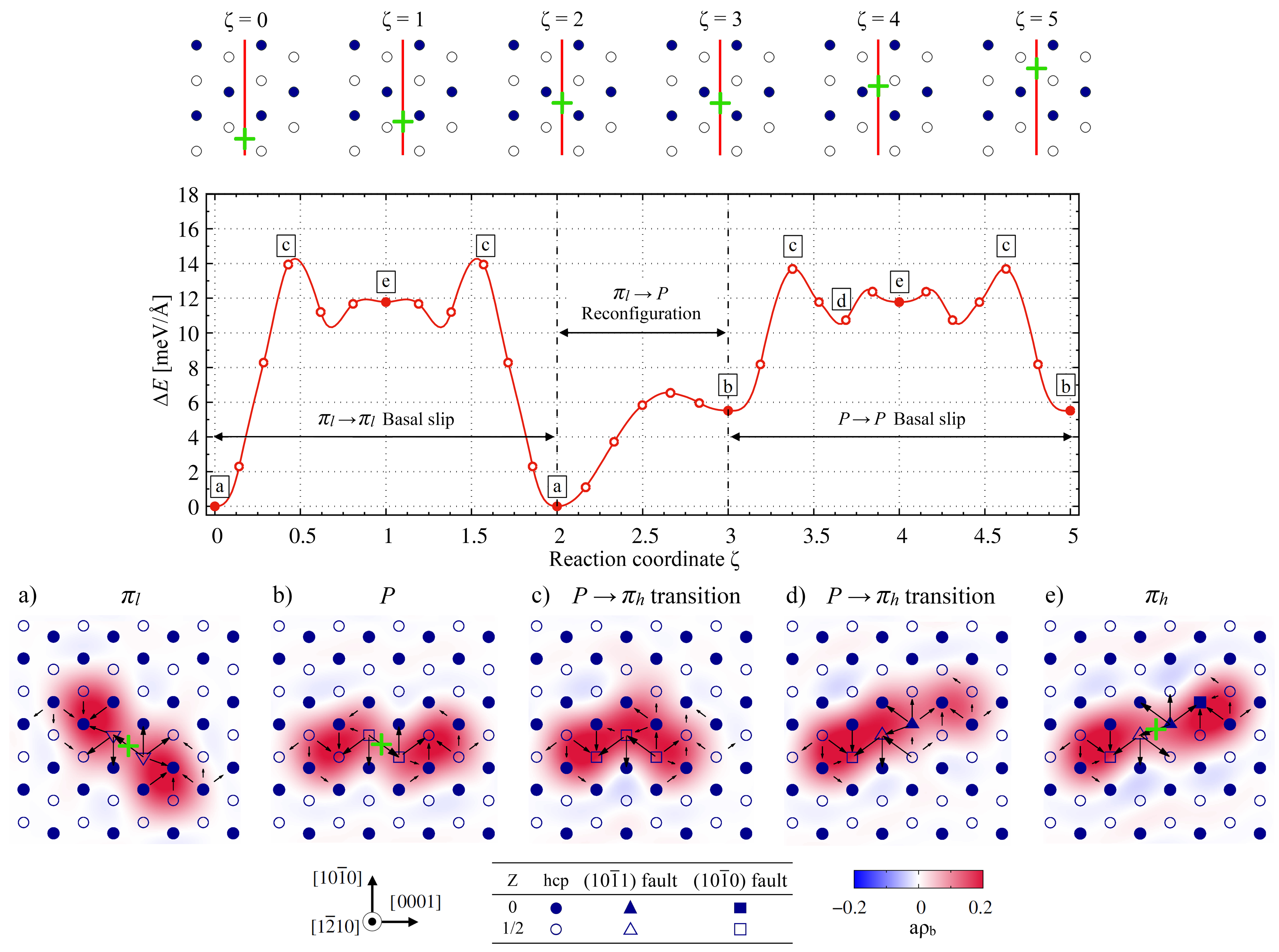}
\caption{Basal slip of the $\hkl<a>$ screw dislocation
starting from its ground-state pyramidal $\pi_l$ and its prismatic $P$ configurations.
The energy variation $\Delta E$ is shown as a function of a reaction coordinate $\zeta$,
with fully colored points and opened circles corresponding to (meta-)stable cores
and transition configurations obtained by the NEB method, respectively.
The corresponding path is sketched on top with the basal slip plane marked by a vertical red line.
Some core structures along this path are shown in subfigures (a-e) 
(see Fig. \ref{Fig1} for a detailed caption of these subfigures).
}
\label{Fig2}
\end{figure*}

The dislocation initial configuration is dissociated in the basal plane.  
It is introduced in the simulation cell by applying to all atoms the elastic displacement predicted 
by anisotropic elasticity for a pair of $1/3\,\hkl<1-100>$ and $1/3\,\hkl<0-110>$ Shockley partials 
in the same prismatic plane (Fig. \ref{Fig1}a), 
thus creating a stacking fault ribbon between them.
Two core positions, which correspond to the center of symmetry of the ground state or of the high energy pyramidal metastable configuration, marked as \textit{B\textsubscript{1}} and \textit{B\textsubscript{2}}, respectively, are considered. 
Width of the core is characterized by the distance $\lambda$ between the two Shockley partials which is measured in units of the distance $k=a\sqrt{3}/2$ between two Peierls valleys in the \hkl[10-10] direction.
Such prepared basal screw dislocations subjected to \textit{ab initio} ionic relaxation spontaneously reconfigure to the ground state $\pi{}_l$ or high energy pyramidal $\pi{}_h$ geometry depending on their initial position (Fig. \ref{Fig1}b), with an energy difference $\Delta E=11.9$\,meV/{\AA} between these two states in good agreement with previous \abinitio studies \cite{Clouet2015,Kwasniak2017,Poschmann2018}. 
Our investigation therefore conclude to the instability of basal dissociation in pure Ti, which is in line with the \abinitio calculations of Tarrat \etal \cite{Tarrat2014} relying on different boundary conditions.

To unveil the energy variation during decay of the basal dissociation, i.e. to estimate the core energy as function of the basal dissociation width, constrained relaxations have also been performed.
In these calculations, atoms in the dissociation plane located between the two partial dislocations were not allowed to relax along the \hkl[1-210] screw direction. 
These atoms are indicated by light-blue squares in Fig. \ref{Fig1}c.
Such a constraint allows maintaining the basal stacking fault.
Because of hcp lattice symmetry, the dissociation distance $\lambda$ where this constraint is applied needs to be equal to an odd (respectively even)
number of distances $k$ between Peierls valleys for the $B_1$ (respectively $B_{2}$) position.
Independently from the selected core position and width, the leading and trailing partials are never purely basal but spread onto the prismatic or first-order pyramidal planes (Fig. \ref{Fig1}c).
Moreover, the energy of the considered line defects strongly depends on the basal dissociation distance as shown in Fig. \ref{Fig1}d.
The excess energy $\Delta E$ continuously increases with the dissociation distance $\lambda$, 
without showing any minimum at small distances, thus indicating the absence of a metastable configuration dissociated in the basal plane. 
The energy variation corresponding to a basal dissociation was also calculated according to elasticity theory (Eq. 3 in Ref. \cite{Clouet2012}) using a core radius equal to 1\,$k$, a stacking fault energy $\gamma_{\rm b}=306$\,mJ/m$^2$ and the elastic constants given by our \abinitio calculations. 
As shown in Fig.\,\ref{Fig1}d, linear elasticity, although not fully quantitative, provides a reasonable estimate of $\Delta E(\lambda)$: 
this energy variation therefore mainly results from the increase of the dissociation width.
Elasticity theory predicts a dissociation distance $d_{\rm b}^{\rm eq}=2.6\,\mathrm{\AA}=1.03\,k$,
a distance which is too small to lead to any real basal dissociation, in agreement with \abinitio results.

\begin{figure}[ht]
\centering
\includegraphics[width=\columnwidth]{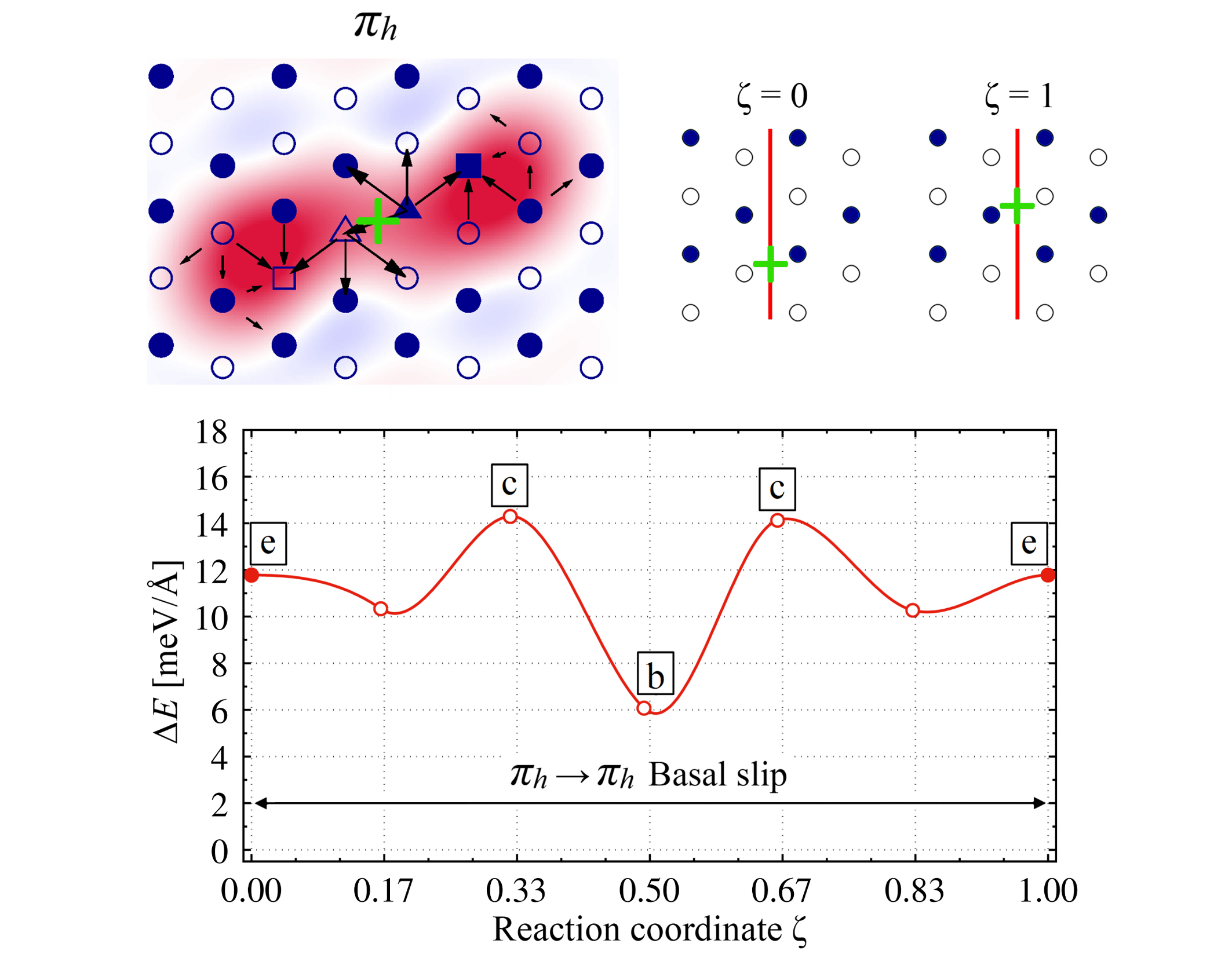}
\caption{Minimum energy path for basal slip of $\hkl<a>$ screw dislocation,
corresponding to the transition between two neighboring $\pi_h$ metastable  core (shown in the top-left corner).
The labels (b), (c), and (e) refer to the configurations shown in Fig. \ref{Fig2}.
}
\label{Fig3}
\end{figure}

As basal slip in pure Ti does not stem from a basal dissociation, 
we consider now the ground state of the screw dislocation, 
where the core is dissociated in a pyramidal plane
(configuration $\pi_l$ in Fig. \ref{Fig1}b),
and examine how such a core can glide in a basal plane.
The studied scenario involves basal motion of the screw dislocation 
from its $\pi{}_l$ ground state to the next equivalent $\pi{}_l$ position in the \hkl[10-10] direction.
The minimum energy path determined with the nudged elastic band (NEB) method \cite{Sheppard2008}
using 5 intermediate images
corresponds to the reaction coordinate $0\leq\zeta\leq2$ in Fig. \ref{Fig2}. 
According to the preliminary results obtained with a force convergence criterion of 30\,meV/\AA,
direct $\pi{}_l$\,\textrightarrow\,$\pi{}_l$ glide goes through an intermediate metastable configuration which corresponds to the high energy $\pi_h$ core (Fig. \ref{Fig2}e).
The NEB calculation was thus further refined by considering the two half-paths $\pi{}_l \rightarrow \pi{}_h$ and $\pi{}_h \rightarrow \pi{}_l$
with 5 intermediate images and a 20\,meV/{\AA} convergence criterion.
The obtained energy barrier is only about +2\,meV/\AA{} higher than the metastable $\pi{}_h$ core and corresponds to a configuration where the dislocation is spread in both prismatic and pyramidal planes (Fig. \ref{Fig2}c).

We then consider the prismatic configuration of the screw dislocation.     
As previously shown \cite{Clouet2015}, $\hkl<a>$ screw dislocations in Ti
can adopt a configuration where the core is fully dissociated in a prismatic plane
and where the dislocation center is at the same position as for the ground state
(Fig. \ref{Fig2}b).
This metastable $P$ core has an energy $\Delta E=5.6$\,mev/\AA,
still in good agreement with previous \abinitio calculations 
\cite{Clouet2015,Kwasniak2017,Poschmann2018},
and the energy barrier leading to this stationary core reconstruction 
is only slightly higher than the energy of $P$ state
(energy path shown for $2\leq\zeta\leq3$ in Fig. \ref{Fig2}).
The minimum energy path for basal glide of this prismatic core geometry
($3\leq\zeta\leq5$ in Fig. \ref{Fig2}) goes through the same configurations
as for the $\pi_l \rightarrow \pi_l$ transition, thus leading to the same 
energy barrier. 
Similarly to $\pi_l \rightarrow \pi_l$ slip, 
after an initial NEB calculation with 5 intermediate images,where the intermediate $\pi_h$ configuration appeared, the $P \rightarrow P$ transition has been refined by considering half the path, corresponding thus to the $P \rightarrow \pi_h$ transition.
Besides the intermediate $\pi_h$ configuration, a well defined additional metastable state (Fig. \ref{Fig2}d) appears
both during $P\rightarrow P$ basal slip ($\zeta \sim 3.7$ and 4.3), 
and during $\pi_l \rightarrow \pi_l$ basal slip ($\zeta \sim 0.7$ and 1.3),
where the dislocation is partly dissociated in the pyramidal and prismatic plane.

We finally consider basal slip of the metastable $\pi_h$ core (Fig. \ref{Fig3}),
using 5 intermediate images in the NEB calculation,
as this configuration appears in both $\pi_l \rightarrow \pi_l$ and $P \rightarrow P$ transition paths.
No new path was actually obtained as this $\pi_h \rightarrow \pi_h$ transition 
goes through the prismatic configuration and is equivalent to the $P \rightarrow P$ basal slip.

Summarizing the present study, it was found that basal glide of the 1/3\textlangle{}11$\bar{2}$0\textrangle{} screw dislocations in pure Ti is realized 
without any basal dissociation.
Instead, the dislocation remains dissociated in pyramidal and prismatic planes during the transition,
with a motion of the partial dislocations and of the stacking fault ribbon in a direction perpendicular to the dissociation plane.
This basal slip mechanism is similar to the one already identified in Zr \cite{Chaari2014}.
Because of the high energy barrier associated with such a path, 
basal slip will proceed through the nucleation of double kinks thanks to thermal activation, 
in agreement with the Peierls mechanism inferred from TEM observations \cite{Caillard2018}.
The same configurations being involved for basal as for prismatic and pyramidal motion, 
this explains the wavy slip traces experimentally observed 
when basal slip is active \cite{Caillard2018,Barkia2017}.
Finally, as the Peierls barrier of basal and pyramidal \citep{Clouet2015} glide are close, both basal and pyramidal slip are activated at high enough temperature \cite{Caillard2018,Barkia2017}.

\textbf{Acknowledgments} -
This work supported by the National Science Center under SONATINA 1 project No. 2017/24/C/ST8/00123. Computing resources were provided by the HPC facilities of the PL-GRID and CI-TASK infrastructure. 

\section*{References}
\bibliography{ref}
\biboptions{sort&compress}

\end{document}